\documentclass[12pt,preprint]{aastex}

\input{epsf}

\newcommand{\msun}{{\rm M}_{\odot}}

\newcommand{\gsim}{\mathrel{\hbox{\rlap{\hbox{\lower4pt\hbox{$\sim$}}}\hbox{$>$}}}}
\newcommand{\lsim}{\mathrel{\hbox{\rlap{\hbox{\lower4pt\hbox{$\sim$}}}\hbox{$<$}}}}

\shortauthors{Hoyle, Rojas, Vogeley \& Brinkmann} 
\shorttitle{The Luminosity Function of Void Galaxies}

\begin{document}

\title{The Luminosity Function of Void Galaxies in the Sloan Digital Sky Survey}

\author{Fiona Hoyle$^1$, Randall R. Rojas$^1$, Michael S. Vogeley$^1$ \& John Brinkmann$^2$ \\
1. Department of Physics, Drexel University, 3141 Chestnut Street,
Philadelphia, PA 19104 \\ 
2. Apache Point Observatory, P. O. Box 59, Sunspot, NM 88349-0059 \\
\email{hoyle@venus.physics.drexel.edu,
rrojas@mercury.physics.drexel.edu, vogeley@drexel.edu} }

\begin{abstract}

The Sloan Digital Sky Survey now extends over a large enough region
that galaxies in low density environments can be detected. Rojas et
al. have developed a technique to extract more than 1000 void galaxies
with $\delta \rho/\rho < -0.6$ on scales r $> 7 h^{-1}$Mpc from the
SDSS. In this paper, we present the luminosity function of these
galaxies and compare it to that of galaxies at higher densities.  We
obtain a measurement of the void galaxy luminosity function over the
range $-21.5<$ M$_{\rm r}^*<-14.5$ in the SDSS r-band. The void galaxy
luminosity function (LF) is well fit by a Schechter function with
values of $\Phi^*=(0.19\pm0.04) \times 0.01 h^3$Mpc$^{-3}$, M$_{\rm
r}^*$ - 5log$h=-19.74\pm0.11$ and $\alpha=-1.18\pm0.13$, where as the
LF of the wall galaxy has Schechter function parameters
$\Phi^*=(1.42\pm0.03) \times 0.01 h^3$Mpc$^{-3}$, M$_{\rm r}^*$ -
5log$h=-20.62\pm0.08$ and $\alpha=-1.19\pm0.07$. The fainter value of
M$_{\rm r}^*$ is consistent with the work of Rojas et al. who found
that void galaxies were fainter than galaxies at higher densities but
the value of $\alpha$ is very similar to that of galaxies at higher
densities. To investigate this further, the LF of wall and void
galaxies as a function of density is considered. Only the wall
galaxies in the highest density regions have a shallower faint end
slope. The LF of the void galaxies is also compared to the LF of the
wall galaxies with and without similar properties to the void galaxies
(color, Sersic index and strength of H$\alpha$ emission). The LF of
the bluest, Sersic index $n<2$ and EW(H$\alpha)>5\AA$ wall galaxies is
very similar in shape to that of the void galaxies, with only slightly
brighter M$_{\rm r}^*$ values and similar values of $\alpha$. The LFs
of the red, high Sersic index, low EW(H$\alpha$) wall galaxies have a
much shallower slope and brighter M$_{\rm r}^*$ values. The value of
$\alpha \approx -1.2$ found from the void galaxy LF  
is shallower that that found from CDM models,
suggesting that voids are not filled with a large population of dwarf
galaxies.

\end{abstract}


\section{Introduction}

One of the most fundamental properties of a distribution of galaxies
is the luminosity function (LF) where $\Phi(L)= dN/dLdV$.
Measurements of the galaxy LF have improved over recent years due to
the large size of new redshift surveys, such as the 2dF Galaxy
Redshift Survey (2dFGRS) and the Sloan Digital Sky Survey (SDSS). For
example, Norberg et al. (2002b) measured the $b_J$-band LF from a
sample of 110,500 galaxies from the 2dFGRS and Blanton et al. (2002)
measured the r-band LF from a sample of 147,986 galaxies taken from
SDSS.

The samples of galaxies are now large enough that they can be split
into subsamples and the properties of different types of galaxies can
be studied in more detail. Projects based on 2dFGRS data include
Madgwick et al. (2002) who measured the LF as a function of spectral
type, Norberg et al. (2002a) who studied the dependence of galaxy
clustering strength on luminosity, and Lewis et al. (2002) who studied
the environmental dependence of galaxy star formation rates. In the
case of the SDSS, Nakamura et al. (2003) have studied the LF as a
function of galaxy morphology, Hogg et al. (2002, 2003) have studied
the overdensities of galaxy environments as a function of luminosity
and color, G\'omez et al. (2002) have studied galaxy star-formation as a
function of environment and Goto et al. (2003) studied the environment
of passive spiral galaxies.

There are many measurements of the LF in many different wavebands and
over a wide range of environments. We restrict our discussion to LFs
measured from optical surveys that cover wide areas. See de Lapparent
(2003) for a comprehensive review of LFs. Most of these analyses have
focused on the LF of galaxies found in regions with densities equal or
higher to the density around field galaxies (e.g. Loveday et al. 1992;
Marzke et al. 1994a, b; Zucca et al. 1997; Ratcliffe et al. 1998; Lin
et al. 1996; Folkes et al. 1999; Blanton et al. 2001, 2002; Norberg et
al. 2002b). In this paper, we will focus on the LF of galaxies found
in extremely {\it underdense} environments ($\delta \rho/\rho < -0.6$
measured on scales of $7h^{-1}$Mpc), which we will refer to as void
galaxies. 

The reason we wish to study the LF of void galaxies is to shed light
on the discrepancies between theory and observation. CDM models
predict the existence of low mass halos in voids (Dekel \& Silk 1986;
Hoffman, Silk \& Wyse 1992). If these halos contain dwarf galaxies
then the void galaxy luminosity function should have a steep faint end
slope. However, surveys of dwarf galaxies indicate that they trace the
same overall structures as `normal' galaxies (Bingelli 1989) and
pointed observations toward void regions have also failed to detect a
significant population of faint galaxies (Kuhn, Hopp \& Els\"{a}sser
1997; Popescu, Hopp \& Els\"{a}sser 1997). Also, including the effects
of photoionization in theoretical models has been shown to restrict
galaxy formation, which leads to a LF with a flatter faint-end slope
(Benson et al. 2003). However, observationally the shape of the void
galaxy LF remains unknown. Grogin \& Geller (1999) measured the void
galaxy LF for 46 galaxies in regions with density less than half of
the mean density (i.e. $\delta \rho/\rho < -0.5$). They found that the
LF of void galaxies was quite steep, $\alpha = -1.4 \pm 0.5$. This
value lies between the steep values predicted by CDM and the shallower
values found observationally for galaxies in denser environments.

In this paper we have a sample of $10^3$ void galaxies selected from
the SDSS from which the void galaxy LF can be measured. Rojas et
al. (2003a, b) has already studied the properties of these
galaxies. They find that void galaxies are fainter, bluer, have
surface brightness profiles more similar to late-type galaxies and
that they have higher specific star formation rates. Goldberg et
al. (2003 in prep) have also investigated the form of the mass
function in voids. They find that the observed mass function is
consistent in both shape and normalization with a semi-analytical mass
function in an underdense environment with $\delta \rho/\rho < -0.5$.
This paper is outlined as follows. In section \ref{sec:survey} we
describe the SDSS and how void galaxies are selected from it. In
section \ref{sec:meth} we describe how the LF is estimated and present
the results in section \ref{sec:res}. In section \ref{sec:lftype} we
consider the LF of different sub-samples of the data, in section 
\ref{sec:discuss} we interpret the results and compare to other
studies and in section \ref{sec:conc} we give conclusions.

\section{The Void Galaxy Sample}
\label{sec:survey}

To obtain a sample of 10$^3$ void galaxies, we use data from the Sloan
Digital Sky Survey. The SDSS is a wide-field photometric and
spectroscopic survey. The completed survey will cover approximately
$10^{4}$ square degrees.  CCD imaging of 10$^8$ galaxies in five
colors and follow-up spectroscopy of 10$^6$ galaxies with $r<17.77$
will be obtained. York et al. (2000) provides an overview of the SDSS
and Stoughton et al. (2002) describes the early data release (EDR) and
details about the photometric and spectroscopic measurements and
Abazajian et al. (2003) describes the first data release (DR1).
Technical articles providing details of the SDSS include descriptions
of the photometric camera (Gunn et al. 1998), photometric analysis (Lupton et
al. 2002), the photometric system (Fukugita et al. 1996; Smith et
al. 2002), the photometric monitor (Hogg et al. 2001), astrometric
calibration (Pier et al. 2003), selection of the galaxy spectroscopic
samples (Strauss et al. 2002; Eisenstein et al. 2001), and
spectroscopic tiling (Blanton et al. 2003a).  A thorough analysis of
possible systematic uncertainties in the galaxy samples is described
in Scranton et al. (2002). All the galaxies are k-corrected and
evolution corrected according to Blanton et al. (2003b) and we assume
a $\Omega_{\rm m}, \Omega_{\Lambda}$ = 0.3, 0.7 cosmology and Hubble's
constant, $h = H_o/ 100$ km s$^{-1}$Mpc$^{-1}$ throughout. 

\begin{figure} 
\begin{centering}
\begin{tabular}{cc}
{\epsfxsize=8truecm \epsfysize=8truecm \epsfbox[40 120 600 700]{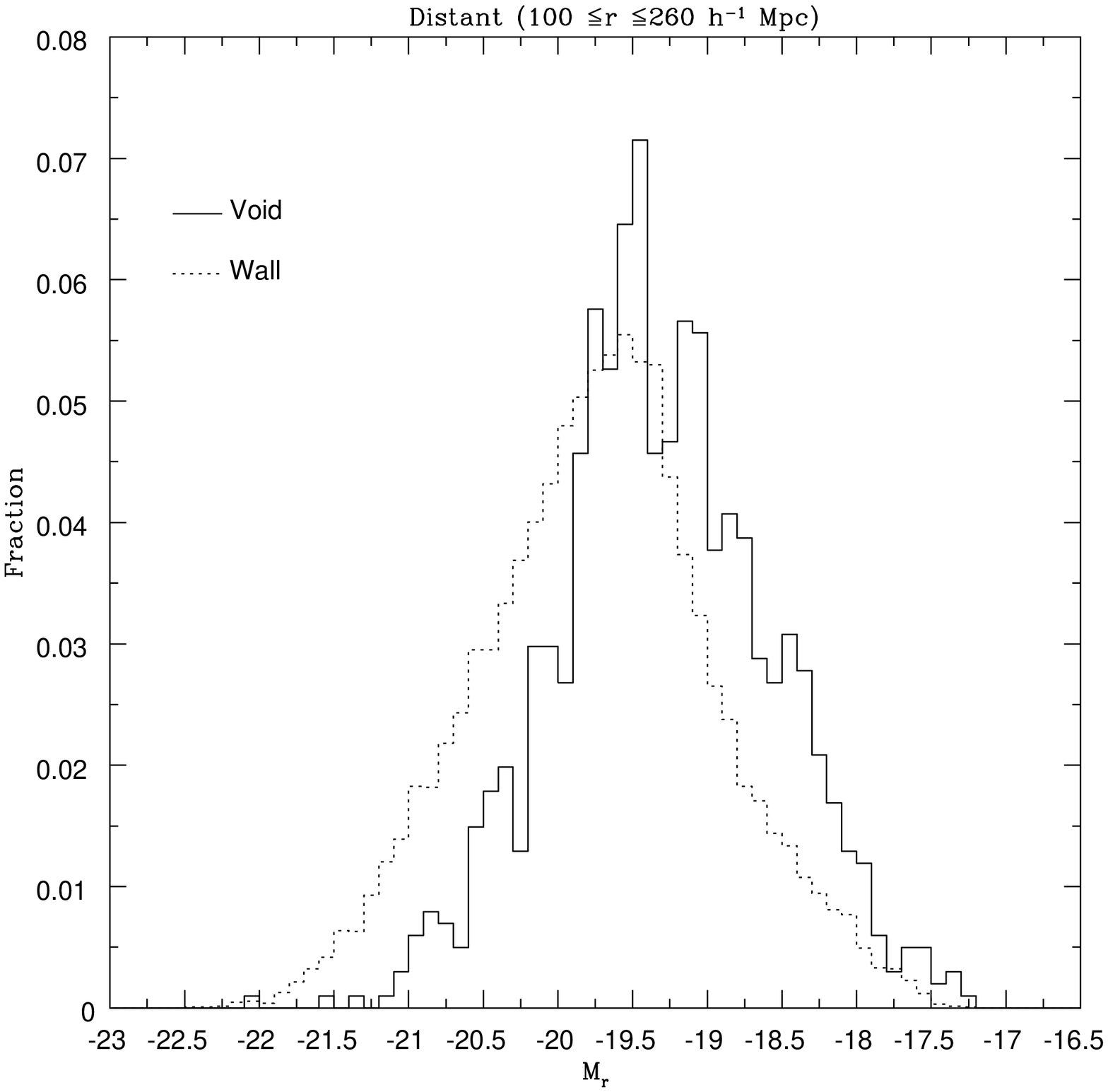}} &
{\epsfxsize=8truecm \epsfysize=8truecm \epsfbox[40 120 600 700]{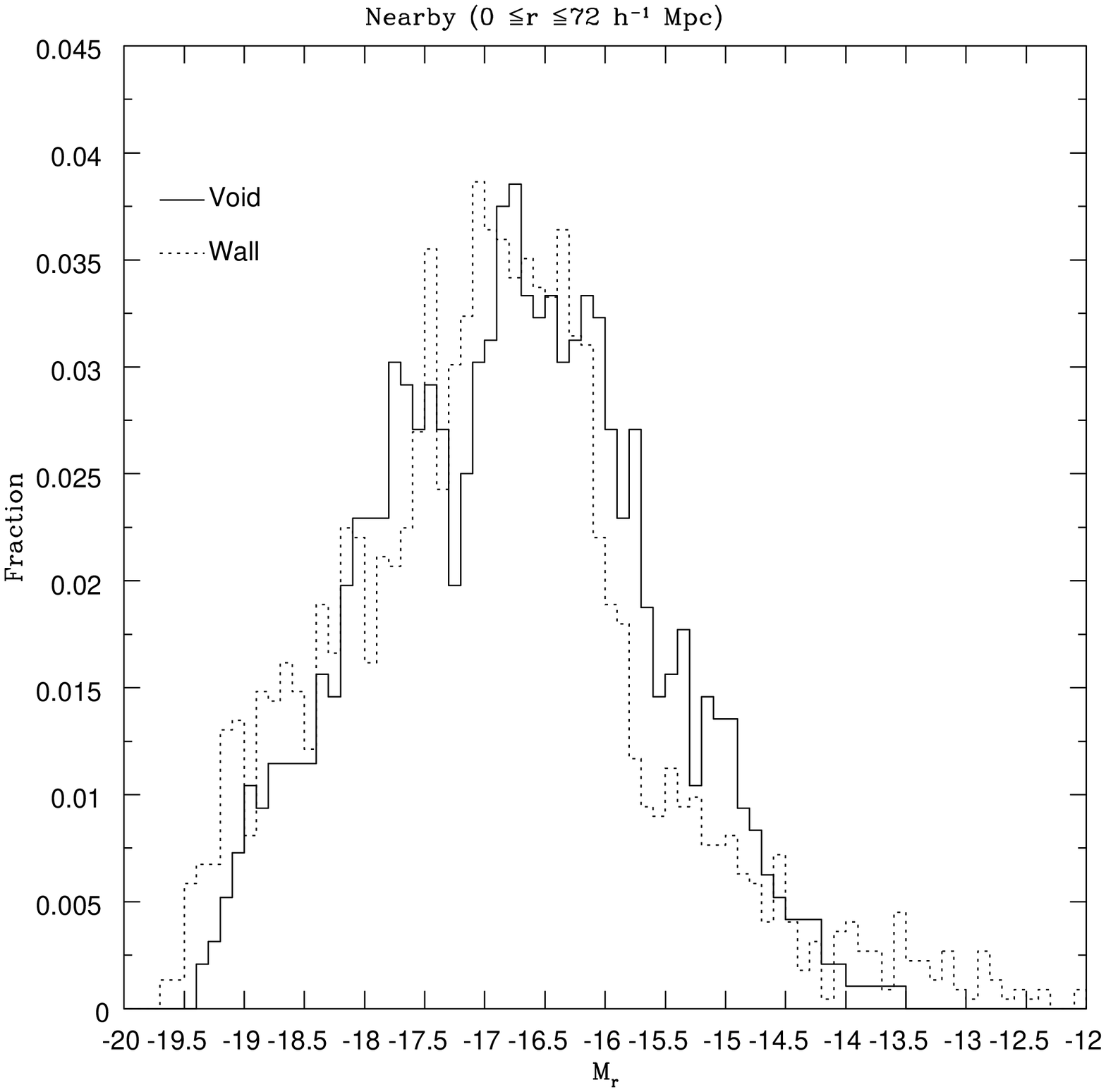}} \\
\end{tabular}
\caption{Distribution of the absolute magnitudes in the distant (left
hand plot) and nearby (right hand plot) wall (dotted line) and void
(solid line) galaxy samples. The range of absolute magnitudes probed
by the distant void galaxies is $-22.0\lsim M_{\rm r} \lsim-17.77$,
where as the nearby void galaxies cover the magnitude range
$-19.7\lsim M_{\rm r} \lsim-13.0$. There are 1,010 void galaxies and
12,732 wall galaxies in the distant sample and 194 void galaxies and
2,256 wall galaxies in the nearby sample.}
\label{fig:hist}
\end{centering}
\end{figure}

Void galaxies are drawn from a sample referred to as {\tt sample10}
(Blanton et al. 2002). This sample covers nearly 2000 deg$^2$ and
contains 155,126 galaxies. We use a nearest neighbor analysis to
select galaxies that on scales of 7$h^{-1}$Mpc have density contrast,
$\delta \rho / \rho < -0.6$ as void galaxies. In the future we wish to
identify void galaxies with even lower values of $\delta \rho / \rho$
but with this sample it is not possible. The choice of $\delta \rho /
\rho < -0.6$ for the density around void galaxies is also consistent
with other definitions of voids. For example the {\tt voidfinder} of
Hoyle \& Vogeley (2002) identifies on scales of 7$h^{-1}$Mpc whether
galaxies are `void' or `wall' galaxies. The void galaxies can reside
in voids reqions but even though 10\% of galaxies are classified as
void galaxies, the void regions still have average values of $\delta
\rho / \rho < -0.9$. Other techniques such as the method of El-Ad \&
Piran (1997) or use of tessellation techniques such as the Delaunay
Tessellation described in Hoyle et al. (2003 in prep.) could also be
used to find void galaxies but currently the geometry of the SDSS does
not allow these techniques to be used as they require large contiguous
volumes. The exact process of selecting the void galaxies is described
in detail in Rojas et al. (2003a). We provide a brief overview, as
follows:

First, a volume limited sample with z$_{\rm max}=0.089$ is
constructed.  This is used to trace the distribution of the voids. Any
galaxy in the full flux-limited sample with redshift z$<$z$_{\rm max}$
that has less then three volume-limited neighbors in a sphere with
radius 7$h^{-1}$Mpc and which does not lie close to the edge of the
survey is considered a void galaxy. Galaxies with more than 3
neighbors are called wall galaxies.  Flux-limited galaxies that lie
close to the survey boundary are removed from either sample as it is
impossible to tell if a galaxy is a void galaxy or if its neighbors
have not yet been observed. This produces a sample of 1,010 void
galaxies and 12,732 wall galaxies. These void and wall galaxies have
redshifts in the range $0.034\lsim z < 0.089$ and, more importantly
for the LF, the void galaxies have magnitudes in the range
$-22<$M$_{\rm r}<-17.77$, which we show in the left hand plot of
figure \ref{fig:hist}. This sample is referred to as the distant
sample.

The SDSS observing schedule is to observe narrow (2.5 degree) but long
($\sim 100$ degree) great circles on the sky. Therefore, there is
little nearby volume in the SDSS and all the galaxies lie close to the
SDSS boundary. However, the wider angle Updated Zwicky Catalog (Falco
et al. 1999) and the Southern Sky Redshift Survey (da Costa et
al. 1998) can be used to trace the distribution of voids locally (see
figure 1. in Rojas et al. 2003a). Volume-limited samples with
approximately the same density as the SDSS volume-limited sample are
constructed. These have z$_{\rm max}=0.025$. We apply the same nearest
neighbor analysis and identify a further 194 void galaxies and 2,256
wall galaxies. The absolute magnitude of the galaxies is
$-19.7<$M$_{\rm r}<-13$, shown in the right hand plot
of figure \ref{fig:hist}. This sample is referred to as the nearby
sample.

The properties of the galaxies that we are interested in are there
magnitudes, colors, surface brightness profiles and star formation
rates. The magnitudes that are measured from the SDSS are petrosian
magnitudes (Petrosian 1976). More details are given in Stoughton et
al. (2002) but basically the petrosian magnitude is the total amount
of flux within a circular aperture whose radius depends on the shape
of the galaxy light profile, i.e., the size of the aperture is not
fixed so galaxies of the same type are observed out to the same
physical distance at all redshifts. The colors of the galaxies are
found using petrosian magnitudes in two bands. We use the g-r color as
the u-band data can be noisy. To estimate the form of the surface
brightness profile we use the Sersic index (Sersic 1968) measured by
Blanton et al. (2002). The Sersic index, $n$, is found by fitting the
functional form $I(r)=I_0 {\rm exp}(-r^{1/n})$. A value of $n=1$
corresponds to a purely exponential profiles, while $n=4$ is a de
Vaucouleurs profile.  The final measurement we consider is the star
formation rate, as measured by the strength of the H$\alpha$ emission
line. We could also consider other lines such as H$\beta$, OII,
NII. Rojas et al. (2003b) finds that on average the equivalent widths
of void galaxies are larger than the equivalent widths of wall
galaxies in all of these lines. As we just want to define a high and
low star formation rate sub-samples (see section \ref{sec:lftype}), the
choice of the line over the others is not critical.

\section{Estimating the Luminosity Function}
\label{sec:meth}

\subsection{Method}

Although referred to throughout this paper as the luminosity function,
we actually calculate the magnitude function, $\Phi(M)$, which is
related to the LF through the relation $\Phi(L) dL = \Phi(M) dM$. To
calculate the LF, we use the maximum likelihood approach of
Efstathiou, Ellis \& Peterson (1988, SWML). Similar results were
obtained using the V$_{\rm max}$ method (Schmidt 1968), apart from at
the faint end where the V$_{\rm max}$ method is known to be a poor
estimator due to sample variance.

We measure the LF of the distant and nearby, void and wall galaxies
separately. The distant and nearby LFs are combined using an error
weighted average of the two measurements in the region of overlap.
The LFs of the wall and void galaxy samples are normalized to match
the number per square degree of galaxies that would be predicted from
the full SDSS sample. Approximately 11/12ths of the galaxies are
considered wall galaxies and 1/12th are void galaxies (Rojas et
al. 2003a).

To measure $\Phi^*$, M$_{\rm r}^*$ and $\alpha$ we $\chi^2$-fit the
void and wall LFs to a Schechter function. We $\chi^2$-fit the
Schechter function twice to each LF: The first time, all three
parameters are allowed to vary to obtain an estimate of M$_{\rm
r}^*$. The second time, we $\chi^2$-fit the three parameters again but
only over the range M$_{\rm r}^*$-1 $<$ M$_{\rm r}<$ M$_{\rm
r}^*$+5 which means we are fitting the Schechter function to the same
part of the LF for each sub-sample.  We do this because the wall
galaxy LF can be measured over a wider range of magnitudes than the
void galaxy LF and thus a comparison between the two LFs might not
otherwise be fair. Estimates of $\alpha$ are most sensitive to the
range of magnitudes over which the fits are performed.  Even so, the
values of $\alpha$ found by the restricted range and the full range of
magnitudes available agree within 1$\sigma$.

\subsection{Errors}
\label{sec:ers}

We use two methods to estimate the uncertainties in the LF
measurements. The first method is that of Efstathiou et al. (1988),
which uses the property that the maximum-likelihood estimates of
$\Phi$ are asymptotically normally distributed. An alternative is the
jackknife method (Lupton 1993).  Following Blanton et al. (2001), we
construct 18 sub samples of the galaxies. Each sub-sample excludes a
different 1/18th of the area of the survey. We measure the LF of each
sample and estimate the error via
\begin{equation}
{\rm Var}(x) = \frac{N-1}{N} \sum_{i=1}^N (x - (\bar x))^2.
\end{equation}

\section{Luminosity Functions of Void and Wall Galaxies}
\label{sec:res}

\begin{figure} 
\begin{centering}
\begin{tabular}{cc}
{\epsfxsize=8truecm \epsfysize=8truecm \epsfbox[40 120 600 700]{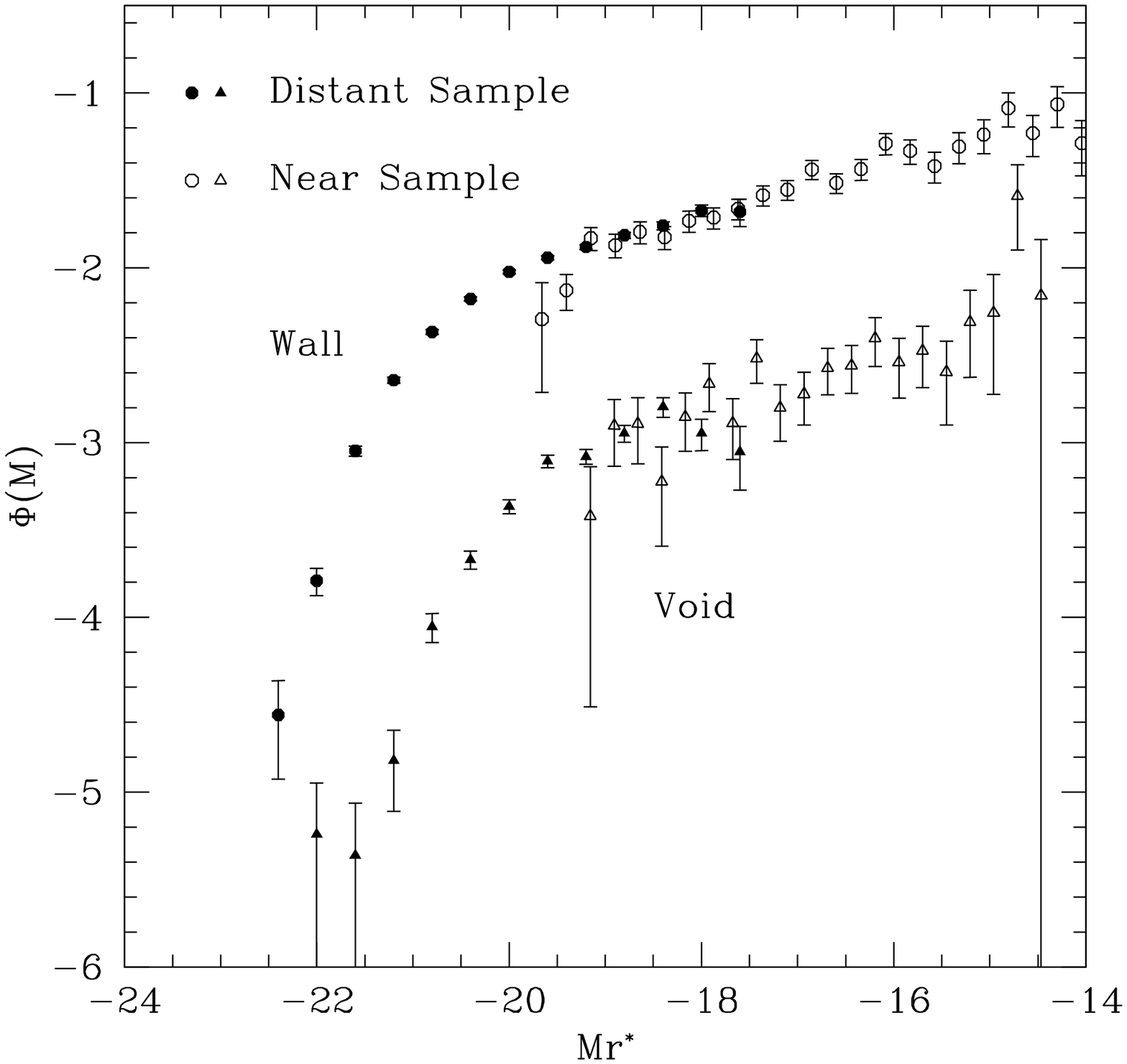}} &
{\epsfxsize=8truecm \epsfysize=8truecm \epsfbox[40 120 600 700]{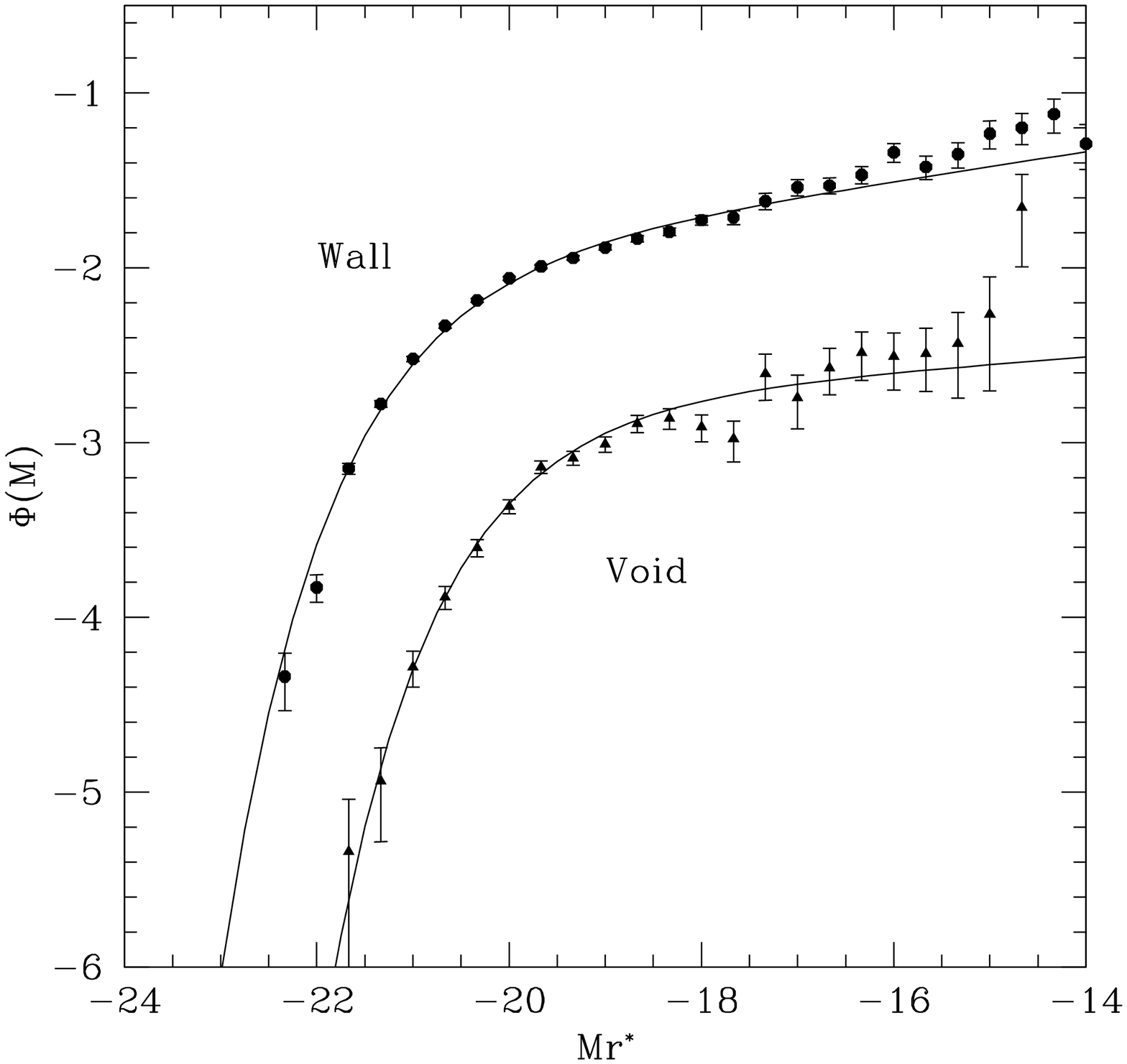}} \\
\end{tabular}
\caption{The Luminosity Function of the different samples of void and
wall galaxies. The left hand plot shows the LF from each of the
distant (solid points) and nearby (open points), void (triangles) and wall
(circles) samples. These samples are combined and the right hand plot
shows the full void (triangle) and wall (circle) LFs. The solid line
shows a best fit Schechter function to each sample. The parameters are
listed in table \ref{tab:values}.}
\label{fig:LFs}
\end{centering}
\end{figure}

In figure \ref{fig:LFs}, we present the LFs of the void and wall
galaxies. The first thing to notice is that we are able to measure the
LF of void galaxies over a wide range of absolute M$_{\rm r}$ values,
from $-21.5<$M$_{\rm r} <-14.5$. This is only possible due to the
large volume of the SDSS and the way in which void galaxies were found
both locally, using the UZC and SSRS2, and at higher redshifts. The
amplitudes of the distant and nearby LFs match well in amplitude in
the region where they sample the same range of absolute magnitude, as
shown in figure \ref{fig:LFs}. In the left hand plot, we show the LFs
estimated from the void (lower amplitude) and wall (higher amplitude),
distant (solid lines) and nearby (dashed lines) samples.  In the right
hand side of figure \ref{fig:LFs}, we show a combined void (circles)
and wall (triangles) LF. The errors shown come from the method of
Efstathiou et al. (1988). The solid lines show best fit Schechter
function curves. These are discussed further below.

\begin{table}
\begin{centering}
\begin{tabular}{lcccc}
Sample & $\Phi^* $  & M$_{\rm r}^*$ - 5log$h$ & $\alpha$ & $\chi^2/ \nu$ \\ 
 & $\times$ 0.01$h^3$Mpc$^{-3}$ & & & \\ \hline
Void Combined  &  0.19$\pm0.04$ & -19.74$\pm0.11$ & -1.18$\pm0.13$  & 0.9 \\
Void Distant  &  0.20$\pm0.08$ & -19.68$\pm0.18$ & -0.97$\pm0.23$  & 1.1 \\
Void Nearby  &   0.31$\pm0.21$ & -20.10$\pm1.90$ & -1.25$\pm0.15$  & 0.5 \\ \hline
Wall Combined  &  1.42$\pm0.3$ & -20.62$\pm0.08$ & -1.19$\pm 0.07$ & 1.9 \\
Wall Distant  &  1.79$\pm0.6$ & -20.40$\pm0.14$ & -1.03$\pm0.16$ & 2.3 \\
Wall Nearby  &  4.20$\pm1.2$ & -18.55$\pm1.29$ & -1.15$\pm 0.12$  & 0.7 \\ \hline
\end{tabular}
\caption{Fitting a Schechter function to the void and wall LFs of the
combined (distant and near LFs combined into one LF), distant, and
near samples. The distant samples alone can constrain $\Phi^* $ and
M$_{\rm r}^*$ well but cannot place strict limits on $\alpha$, whereas the
nearby samples constrain $\alpha$ better. }
\label{tab:values}
\end{centering}
\end{table}

\begin{figure} 
\begin{centering}
\begin{tabular}{c}
{\epsfxsize=8truecm \epsfysize=8truecm \epsfbox[40 120 600 700]{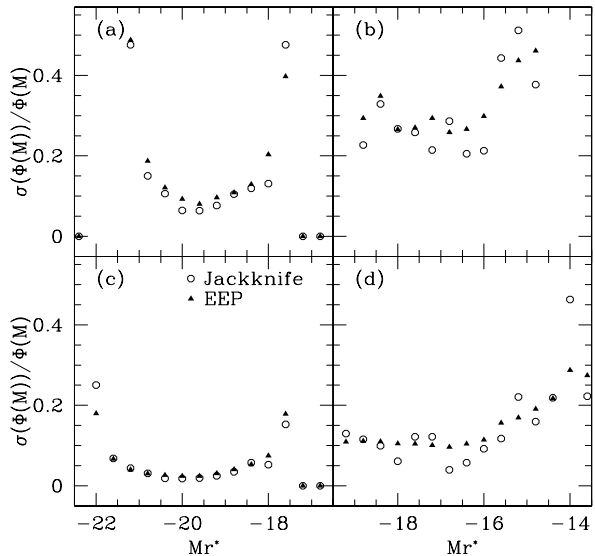}} \\
\end{tabular}
\caption{A comparison of the error estimation techniques. We compare
the fractional error as calculated by the method of Efstathiou et
al. (1988, EEP, triangles) and by the jackknife method (circles),
described in section \ref{sec:ers}.  The numbers in the top left
corner correspond to (a) void distant sample, (b) void nearby sample,
(c) wall distant sample, (d) wall nearby sample. The errors are smaller
for the distant samples due to the larger number of galaxies. Both
methods give similar size errors, although the jackknife errors
exhibit more scatter.}
\label{fig:ers}
\end{centering}
\end{figure}

We compare the errors found via the Efstathiou et al. and jackknife
methods in figure \ref{fig:ers}. Figure \ref{fig:ers} shows the
fractional error as measured by the two methods for each sample,
void/wall, distant/nearby. The errors match each other very well,
especially in places where the LF can be accurately measured. The
jackknife errors tend to vary more from point to point, whereas the
Efstathiou et al. errors follow a smoother trend. As the two methods
give consistent results we will only consider the Efstathiou et
al. errors in any further analysis.

We $\chi^2$-fit a Schechter function to the various LFs as described
in section \ref{sec:meth}. In the case of the void galaxies, the LFs
are very well fit with values of $\chi^2$ per degree of freedom close
to 1. The wall galaxies are less well fit by a Schechter function: the
value of $\chi^2$ per degree is around 2.5. These values probably
reflect the smaller errorbars on the wall galaxy LF more than the fact
that the void galaxies have a LF with a shape that more closely
resembles a Schechter function.

\begin{figure} 
\begin{centering}
\begin{tabular}{c}
{\epsfxsize=8truecm \epsfysize=8truecm \epsfbox[0 0 550 550]{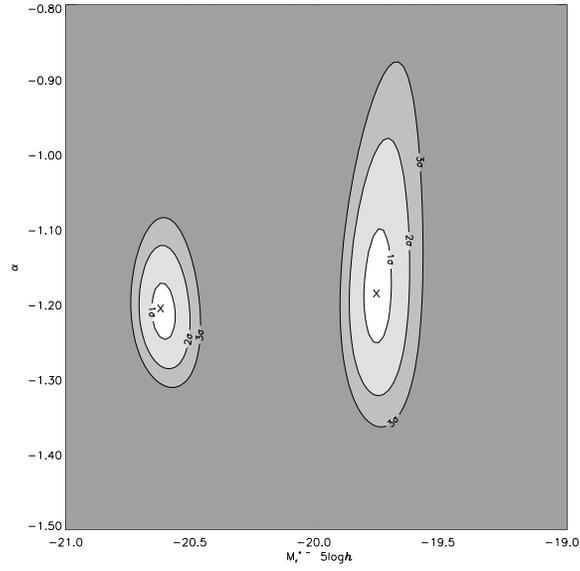}} \\
\end{tabular}
\caption{Comparing the void LFs to a Schechter function. The right
hand contours are for the void galaxies. $\Phi^*$ is fixed to be 0.0019
$h^3$Mpc$^{-3}$, which is the best fit value found allowing $\Phi^*$
also to vary. The contours show the best fit values of M$_{\rm r}^*$ -
5log$h$ and $\alpha$ of 19.74 and -1.18 respectively plus the 1, 2 and
$3-\sigma$ contours. The left hand contours are for the wall
galaxies. $\Phi^*$=0.014 and the best fit values of M$_{\rm r}^*$ - 5log$h$
and $\alpha$ of -20.62 and -1.19 are shown as well as the 1, 2 and
$3-\sigma$ contours.}
\label{fig:cont}
\end{centering}
\end{figure}

The best fit Schechter function parameters to the combined void and
wall galaxy LFs are shown in table \ref{tab:values}. The values from
the combined void and wall samples are used to calculate the Schechter
function curves in figure \ref{fig:LFs}. In figure \ref{fig:cont}, we
fix $\Phi^*$, using the value in table \ref{tab:values}, and present
the contours of $\alpha$ and M$_{\rm r}^*$ found from fitting the
combined LFs. The Schechter function parameters found for the
combined wall sample are similar to those of Blanton et al. (2001),
who obtained $\Phi^*=(1.46\pm0.12) \times 10^{-2} h^3$ Mpc$^{-3}$,
M$_{\rm r}^*$ - 5log$h=-20.83 \pm 0.03$ and $\alpha =-1.20\pm0.03$.
We also fit the distant samples and nearby samples independently. In
the distant sample cases, the value of $\alpha$ is not well
determined. In the nearby sample cases, $\Phi^* $ and M$_{\rm r}^*$
are not well fit. This is because the nearby and distant samples alone
only probe a narrow range of luminosity.  These results are also given
in table \ref{tab:values}.

It may be possible that the LF and the Schechter function parameters
are incorrectly estimated due to the way the nearby and distant
samples are matched together. The value of M$_{\rm r}^*$ and $\alpha$
should not be too affected as only the points around M$_{\rm r}=-17.5$
have little influence on the estimation of these parameters. The
overall normalization is most likely to be affected as the points
around -17.5 might be underestimated. To see what difference the
points in the overlap around M$_{\rm r}$=-17.5 have on the estimation
of the LF, we ignore the points between $-18 <$M$_{\rm r} <-16.7$ and
refit the combined void LF. The Schechter function parameters agree
with those found from fitting the whole LF to within 1$\sigma$. Once
the SDSS is finished it will cover a larger contiguous area, thus the
void galaxies can be detected nearby without the use of the CFA+SSRS2
and the LF estimated from one contiguous sample. The larger number of
void galaxies will also make it possible to consider the bivariate
distribution of surface brightness and luminosity.

\section{Luminosity Function as a function of Density, Color, 
Profile and H$\alpha$ Line Strength}
\label{sec:lftype}

\begin{figure} 
\begin{centering}
\begin{tabular}{c}
{\epsfxsize=8truecm \epsfysize=8truecm \epsfbox[0 0 550 550]{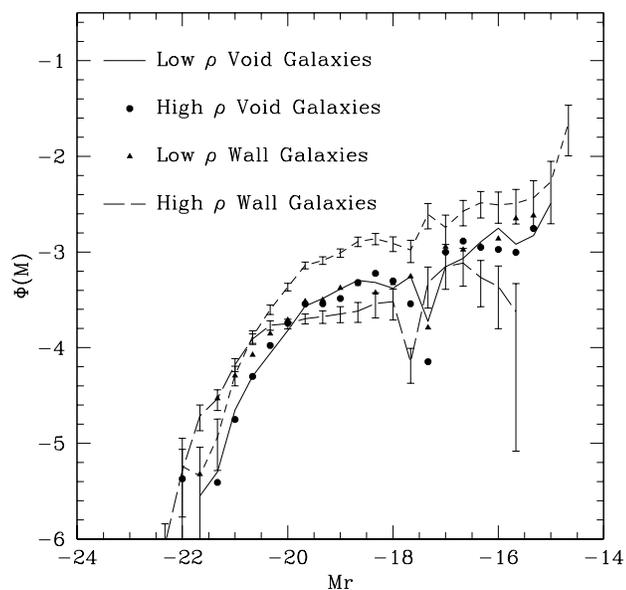}} \\
\end{tabular}
\caption{The LF of galaxies as a function of density. Shown as the
short dashed line is the full void galaxy LF as in figure
\ref{fig:cont}. The solid line shows the LF of the void galaxies in
the most underdense environment $(\delta \rho / \rho < -0.75)$, the
circles - LF of the higher density void galaxies $( -0.75 < \delta
\rho / \rho < -0.6)$, the triangles - LF of the lower density wall
galaxies $( -0.5 < \delta \rho / \rho < 1 )$ and the long dashed line
- LF of the higher density wall galaxies $( 1 < \delta \rho / \rho
)$.}
\label{fig:rho}
\end{centering}
\end{figure}

The sample of void galaxies is large enough that it can be split in
half and the LF measured without the uncertainty in the measurement of
the LF dominating the LF signal. In figure \ref{fig:rho}, the LF of
the void and wall galaxies is considered as a function of local
density. We use the distance to the third nearest neighbor as a
measure of the local density and split both the void and wall galaxy
samples close to the median value, then measure the LF from these
sub-samples.

\begin{table}
\begin{centering}
\begin{tabular}{lcccc}
Sample & $\Phi^* $  & M$_{\rm r}^*$ - 5log$h$ & $\alpha$  & $\chi^2/ \nu$ \\ 
     & $\times$0.01$h^3$Mpc$^{-3}$ &                 &                & \\ \hline
Blue  		   &  0.90$\pm0.5$ & -20.34$\pm0.14$ & -1.32$\pm0.11$ & 1.0 \\
Sersic index $<2$  &  1.30$\pm0.6$ & -20.02$\pm0.16$ & -1.28$\pm0.10$ & 0.9 \\
High EW(H$\alpha$) &  0.90$\pm0.5$ & -19.98$\pm0.17$ & -1.32$\pm0.11$ & 1.0 \\ \hline
Red		   &  0.98$\pm0.5$ & -20.38$\pm0.12$ & -0.73$\pm 0.21$ & 4.8 \\
Sersic index $>2$  &  0.98$\pm0.5$ & -20.34$\pm0.14$ & -0.51$\pm 0.32$ & 8.0 \\
Low EW(H$\alpha$)  &  1.00$\pm0.6$ & -20.36$\pm0.14$ & -0.55$\pm 0.35$ & 7.8 \\ \hline
\end{tabular}
\caption{Fitting a Schechter function to the wall galaxies split by
color, Sersic index and strength of the H$\alpha$ emission. The fits are
made to the LFs shown in figure \ref{fig:split} and described in section \ref{sec:lftype}. }
\label{tab:props}
\end{centering}
\end{table}

The void galaxies all have a distance to the third nearest neighbor
larger than 7$h^{-1}$Mpc which implies a value of $\delta \rho / \rho
< -0.6$ (see Rojas et al. 2003a for details). We split the sample at
8.4$h^{-1}$Mpc which means the lowest density half contains galaxies
with $\delta \rho / \rho < -0.75$ and the higher density half has
galaxies with $ -0.75 < \delta \rho / \rho < -0.6$. The wall galaxy
sample has an median distance to the third nearest neighbor close to
4$h^{-1}$Mpc. The wall galaxies are split here which means the lower
density wall sample has values of $ -0.6 < \delta \rho / \rho < 1 $
and the higher density wall galaxy sample has values of $ 1 < \delta
\rho / \rho $. In this case only, we also sparsely sample the wall
galaxy samples to contain 500 galaxies which is the number of void
galaxies in each sample so that the signal in the void and wall
galaxies is comparable.

We find that the LFs of the high and low density void samples and the
low density wall sample are similar. There is a trend towards brighter
values of M$_{\rm r}^*$ - 5log$h$ ($\sim -19.70, -19.80, -20.20, \pm
0.15$ respectively) with increasing density but the faint end slopes
are quite similar ($\sim -1.15 \pm 0.20$). These LFs are very similar
to the shape of the LF of the full void galaxy sample, but with a
factor of two lower amplitude. The LF of the wall galaxies in the
highest density sample, however, has a somewhat different shape: the
faint end slope, $\alpha$, is shallower ($-0.92\pm 0.20$) and M$_{\rm
r}^*$ - 5log$h$ is brighter ($-20.40\pm 0.15$).

To investigate the shape of the LF further, we would like to study the
void galaxy LF as a function of photometric and spectroscopic
parameters such as color, surface brightness profile (SBP), and line
widths. However, as shown in Rojas et al. (2003a), the void galaxies
only have a narrow range of these properties - they are blue, have
small $n<2$ Sersic indices and have strong equivalent widths (Rojas et
al. 2003b, in prep). Instead, we consider the wall galaxies and split
these into one sample that has properties similar to the void galaxies
and one that does not, i.e. we split on color at g-r=0.75, Sersic
index at $n=2$ and H$\alpha$ equivalent width of 5 \AA, and measure
the LF of each sample of wall galaxies. These values are close to but
not exactly the median value.

\begin{figure} 
\begin{centering}
\begin{tabular}{ccc}
{\epsfxsize=5truecm \epsfysize=5truecm \epsfbox[40 120 600 700]{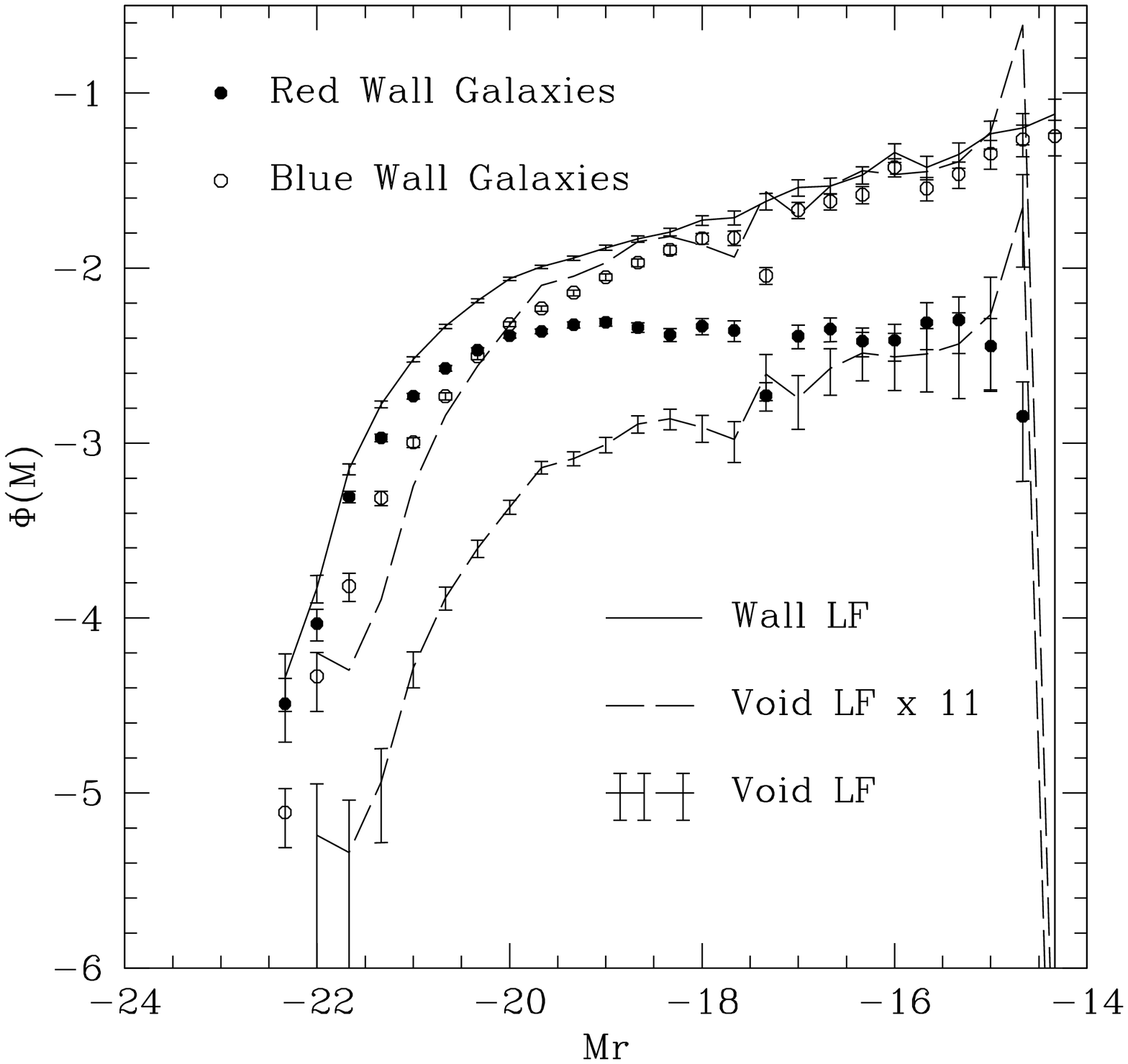}} &
{\epsfxsize=5truecm \epsfysize=5truecm \epsfbox[40 120 600 700]{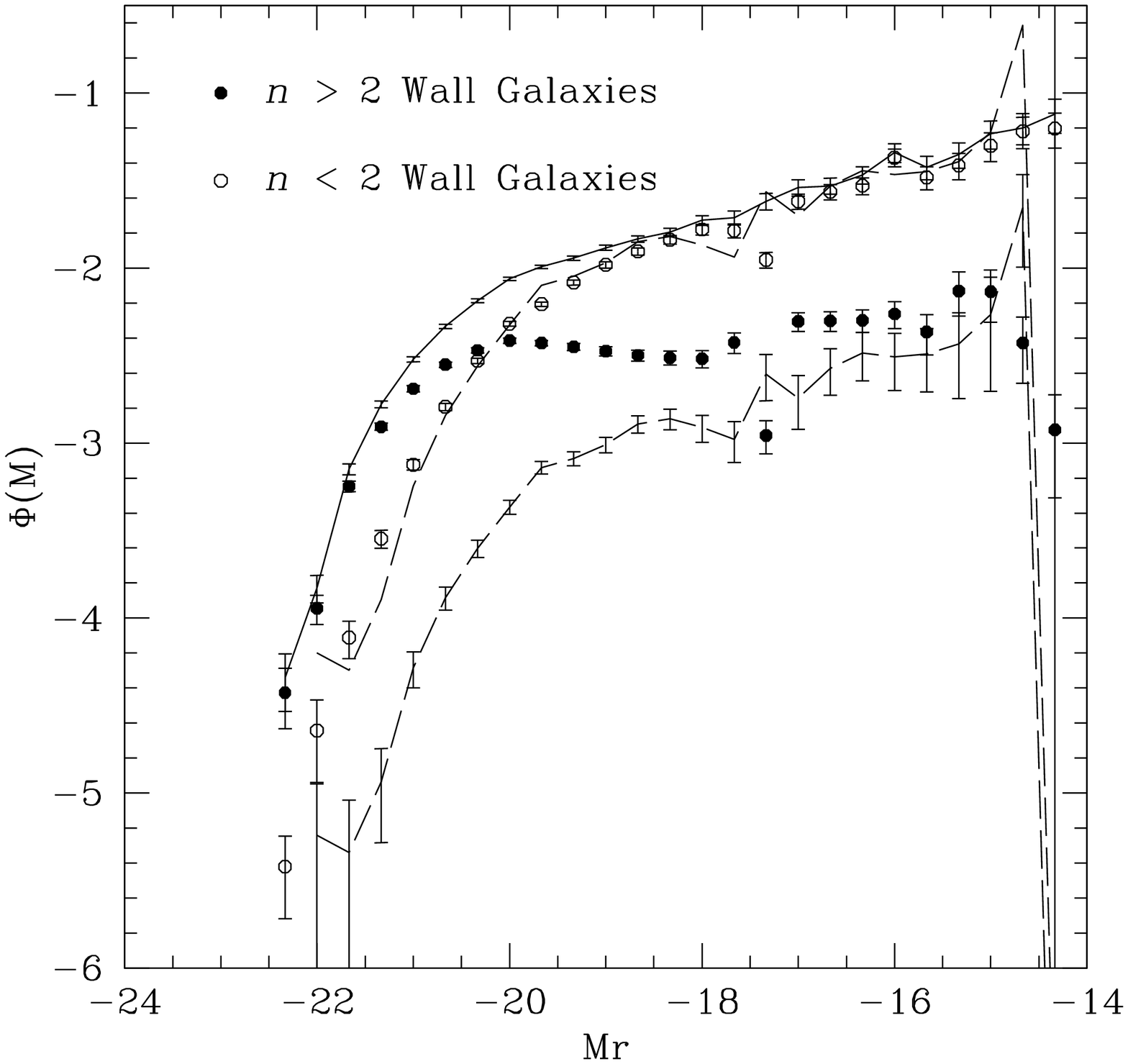}} &
{\epsfxsize=5truecm \epsfysize=5truecm \epsfbox[40 120 600 700]{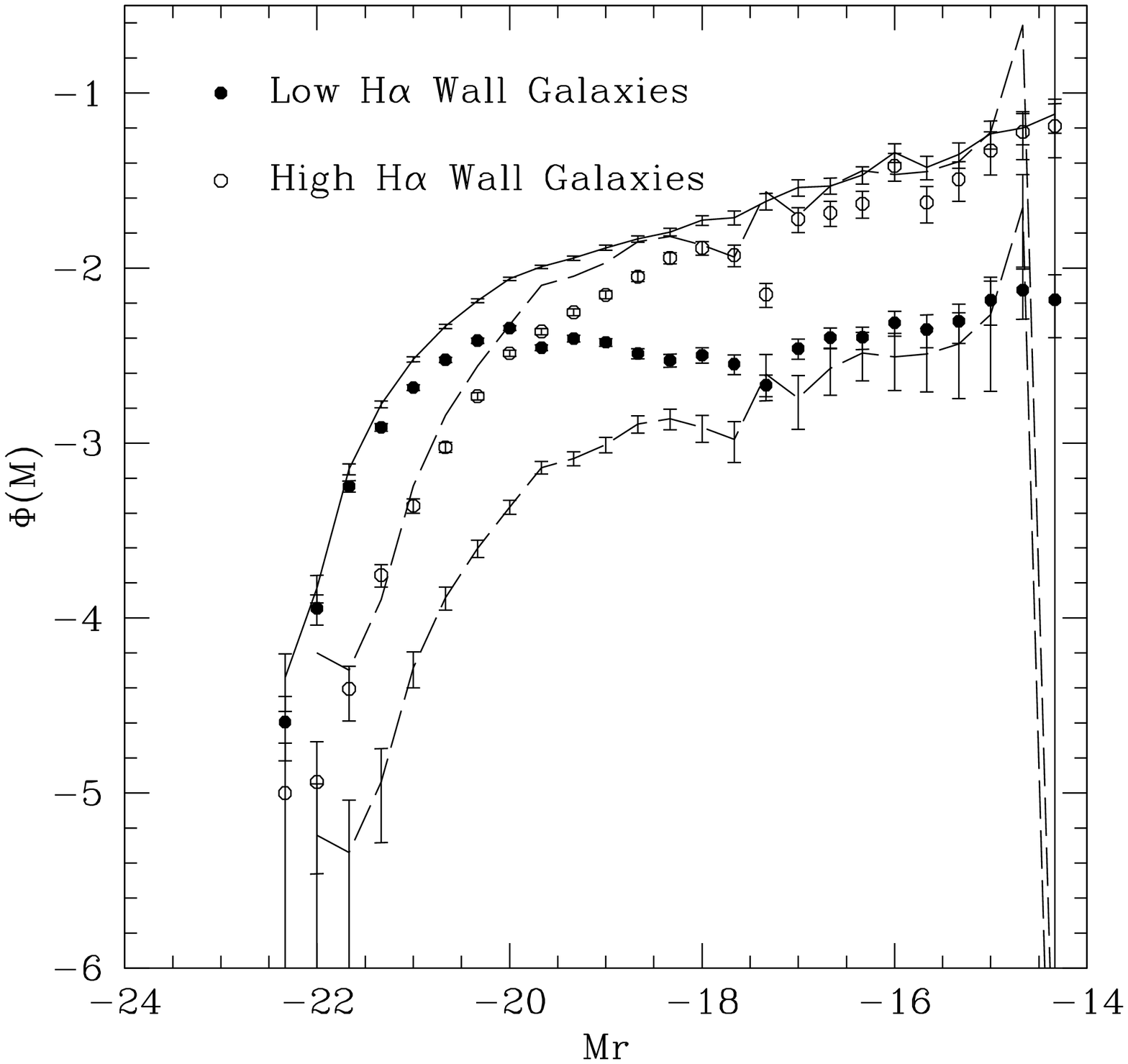}} \\
\end{tabular}
\caption{LF of wall galaxies split by color (left hand plot), Sersic
index (center plot) and strength of H$\alpha$ equivalent width (right
hand plot). The open circles show the LFs of the wall galaxies with
properties that are generally associated with late type galaxies
whereas the filled circles show the LFs of the wall galaxies with
properties that are generally associated with early type galaxies.
The solid line shows the LF of the wall galaxies, the
dashed line with errors shows the LF of the void galaxies and the
higher amplitude dashed line without errorbars shows the LF of the
void galaxies multiplied by 11. }
\label{fig:split}
\end{centering}
\end{figure}

\begin{figure} 
\begin{centering}
\begin{tabular}{c}
{\epsfxsize=6truecm \epsfysize=6truecm  \epsfbox[10 200 600 600]{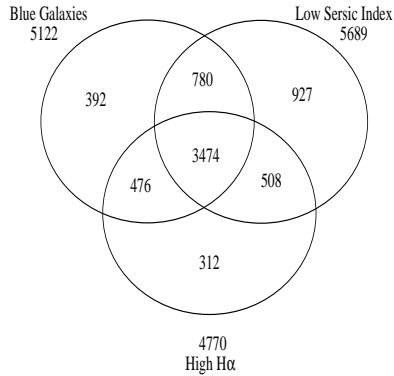}} \\
\end{tabular}
\caption{A venn diagram showing the number of wall galaxies 
that have colors, Sersic indices and H$\alpha$ equivalent widths
similar to the void galaxies. The bulk of the galaxies lie in the
center of the diagram, hence are contained in each sub-sample.}
\label{fig:venn}
\end{centering}
\end{figure}

In each of the panels in figure \ref{fig:split}, we show the LF of the
full void and wall galaxy samples, as shown in figure \ref{fig:LFs},
the LF of the wall galaxies with properties similar to void galaxies
(open circles), and the LF of the wall galaxies with properties
dissimilar to void galaxies (filled circles). We also show the void
galaxy LF multiplied by a factor of 11 to approximately match the
amplitude of the wall galaxy LF as there are $\sim$11 times as many
wall galaxies as void galaxies. We show the blue vs. red wall galaxies
in the left hand plot, the wall galaxies with surface brightness
profiles similar to spiral galaxies vs. elliptical galaxies in the
center plot, and the galaxies with strong H$\alpha$ emission vs. the
galaxies with little H$\alpha$ emission in the right hand plot. In all
cases, the LF of the wall galaxies with void galaxy like properties
appears similar to the LF of the void galaxies whereas the LFs of the
wall galaxies with dissimilar properties to the void galaxies are very
different.  It is not surprising that the LFs of the blue, low Sersic
index and strong H$\alpha$ emission are similar as each sample
contains mostly the same galaxies. In figure \ref{fig:venn} we show
the total number of galaxies in each sample and the number of galaxies
shared between samples. 3,500 of the 5,000 galaxies in each sample are
the same. The same is true for the wall galaxies with dissimilar
properties and the nearby samples.

We again fit a Schechter function to each LF. The blue, low Sersic
index, and high H$\alpha$ LFs all have faint end slopes of around
-1.3$\pm0.1$ and values of M$_{\rm r}^*$ around -20.00$\pm0.15$.  The
best fitting values of $\alpha$ and M$_{\rm r}^*$ for the red, high
sersic and low H$\alpha$ LFs are -0.6$\pm0.2$ and -20.35$\pm0.14$,
although the values of $\chi^2$ per degree of freedom for the red,
$n>2$, H$\alpha<5$ LFs are around 8, meaning that a Schechter
function is a poor fit, although it remains a convenient way of
comparing LFs. We present these number in table \ref{tab:props}.
This shows that the LF of void galaxies is similar to that of
the wall galaxies with similar properties. This might seem in direct
contradiction to Rojas et al. (2003a, b), who found that void galaxies
are bluer, fainter and have diskier SBP's. Our interpretation is that the
void galaxies are just more extreme versions of the late-type wall galaxies,
i.e. the void galaxies are just fainter, bluer and are forming more
stars than the late-type wall galaxies rather than being a completely
different population of galaxies.

\section{Comparison with Previous Observations and Simulations}
\label{sec:discuss}

In this section we compare the results of the void galaxy LF to other
LF measurements from a range of different environments and different
galaxy types. Where possible we compile the measurements of 
M$^*$ and $\alpha$ in table \ref{tab:surveys}.

We find that the LF of the void galaxies has a relatively faint value
of M$_{\rm r}^*$. This shift in M$_{\rm r}^*$ is consistent with the
results of Rojas et al. (2003a) who find that void galaxies are
fainter than wall galaxies. M$_{\rm r}^*$ for the void galaxies is
approximately one magnitude fainter than that of the wall galaxies and
the value measured by Blanton et al. (2002) for the full sample.
However, the faint end slopes of both the wall and void samples are
similar to the value of -1.20$\pm0.03$ obtained by Blanton et
al. (2001) for the SDSS galaxies and -1.21$\pm0.03$ found by Norberg
et al. (2002b) for the 2dFGRS. The wall galaxy LF should be similar to
that of Blanton et al. (2001) but the void galaxy LF could have been
different as the void sample contains only 1/12th of the SDSS galaxies,
after the galaxies close to the survey boundary are removed.

Previous work on the LF of void galaxies has been limited by the small
number of void galaxies available. Grogin \& Geller (1999) present
Schechter function parameter contour plots for their void galaxies.
The LF of the 46 void galaxies with density contrast less than -0.5
has a very steep slope of $\alpha=-1.4$ but the error on this
measurement is $\pm 0.5$. Their full sample of 149 galaxies with
density contrast less than 0 has a slope of $\alpha =-0.5\pm0.3$ in
the B-band and $-0.9\pm0.3$ in the R-band. The void galaxy sample
presented here resides in environments that are even more underdense
than -0.5 but we do not find evidence for such a steep
faint-end slope.

Trentham \& Tully (2002) investigate the faint end of the LF in richer
environments. They study the LF in nearby groups and clusters of
galaxies such as the Virgo Cluster, Coma I Group, Leo Group and two
NGC groups. They find that a Schechter function is a bad fit to the
data as there appears to be a change in the shape of the LF around
M$_R=-18$ which the Schechter function does not account for. This
feature is also seen in our LFs but because the distant and nearby
samples are matched around this magnitude, it is difficult for us to
asses its reality. It will be interesting to see if this shape change
persists when the SDSS is finished and the LFs measured over a wide
range of magnitudes without the need to stitch distant and nearby LFs
together. Trentham \& Tully show that despite the poor fit to a
Schechter function, a faint end slope of $\alpha = -1.19$ is
consistent with all of the groups considered in this study.

As well as looking at the environmental dependence of the LF, we can
consider the morphological dependence of the LF and examine what type
of galaxies have LF shapes similar to the void LF. Nakamura et
al. (2003) looked at morphology dependent r-band LFs of a sample of
1482 SDSS galaxies with magnitudes in the range $-23<M_{\rm r}^*<-18$.
Morphologies of the galaxies were determined by eye. They find that
elliptical galaxies have a shallow faint end slope of $\alpha = -0.83
\pm 0.26$.  Late type spiral galaxies, Sc, d's, also have a shallow
faint end slope of $\alpha = -0.71 \pm 0.26$, but Sa, b's have a slope
of $\alpha=-1.15 \pm0.26$. Only the irregular, Im, galaxies had a
steep faint end slope with $\alpha = -1.9$ with no quoted error as
this was measured from only a hand full of galaxies. This is mostly
consistent with the results of Marzke et al. (1994b) who found that
elliptical, S0, Sa, Sb and Sc, Sd galaxies had b-band faint end slopes
with $\alpha <-1$, but the Sm Im LF had a value of $\alpha =-1.87$,
again determined from a few galaxies. The only inconsistency between
the work of Nakamura et al. and Marzke et al.  is the difference
between the slopes of the Sa, Sb galaxies, which Nakamura found to be
steeper. Measurements in different wave bands may explain this
discrepancy.

\begin{table}
\begin{centering}
\begin{tabular}{llcc}
Reference & Sample &  M$_{\rm r}^*$ - 5log$h$ & $\alpha$  \\  \hline
Blanton et al. (2002) & SDSS         & -20.83$\pm0.03$ & -1.20$\pm0.03$ \\ \hline
Norberg et al. (2002) & 2dFGRS (b$_J$) & -20.65$\pm0.07$ & -1.21$\pm0.03$ \\ \hline
Nakamura et al. (2003) & SDSS E+SO & -20.75$\pm0.17$ & -0.83$\pm0.26$ \\
&	SDSS Sa, Sb & -20.30$\pm0.19$ & -1.15$\pm0.26$ \\
&	SDSS Sb, Sc & -20.30$\pm0.20$ & -0.71$\pm0.26$ \\
& SDSS Im & -20.0 & -1.9 \\ \hline
Nakamura et al. (2003 & SDSS Concentrated & -20.35$\pm0.19$ & -1.12$\pm0.18$ \\ 
& SDSS Diffuse & -20.62$\pm0.14$ & -0.68$\pm0.23$ \\ \hline
Marzke et al. (1994) & CfA E (M$_z$) & -20.34 & -0.85 \\
&	CfA S0 (M$_z$) & -19.85 & -0.94 \\
&	CfA Sa, Sb (M$_z$) & -19.83 & -0.58 \\
&	CfA Sb, Sc (M$_z$) & -19.92 & -0.96 \\
&	 CfA Im (M$_z$) & -19.90 & -1.87 \\ \hline
Grogin \& Geller (1999) & CfA2 + Century (M$_z$) & -20.0$\pm0.2$ & -0.5$\pm0.3$ \\
 & CfA2 + Century (R) & -20.4$\pm0.3$ & -0.9$\pm0.3$ \\ 
 & CfA2 + Century Void (M$_z$) & -20.0 & -1.4$\pm0.5$ \\
 & CfA2 + Century Void (R) & -20.4 & -1.4$\pm0.5$ \\ \hline
Madgwick et al. (2002) & 2dFGRS Very Low SFR (b$_J$) & -20.57$\pm0.05$ & -0.54$\pm0.02$ \\ 
& 2dFGRS Low SFR (b$_J$) & -20.54$\pm0.03$ & -0.99$\pm0.01$ \\ 
& 2dFGRS Moderate SFR (b$_J$) & -20.16$\pm0.04$ & -1.24$\pm0.02$ \\ 
& 2dFGRS High SFR (b$_J$) & -20.14$\pm0.05$ & -1.50$\pm0.03$ \\  \hline
\end{tabular}
\caption{Values of M$_{\rm r}^*$ - 5log$h$ and $\alpha$ found from
other studies. All the values of M$^*$ are shifted into the SDSS
r-band using the following approximations: r - M$_z$ =-1.11, r -
b$_J$= -0.99, which are the shifts appropriate to Sab galaxies
(Fukugita et al. 1995).}
\label{tab:surveys}
\end{centering}
\end{table}

Nakamura et al. (2003) also find that concentrated galaxies have a LF
with a faint end slope of $\alpha = -1.12 \pm 0.18$ whereas less
concentrated galaxies have a faint end slope of $\alpha = -0.68 \pm
0.23$. Rojas et al. (2003a) find that void galaxies have low values of
the concentration index $r_{90}/r_{50}$ hence are concentrated thus
the void galaxy LF is consistent with the LF of the concentrated
galaxies in Nakamura et al. 

Madgwick et al. (2002) measures the LF of the galaxies in the 2dFGRS
as a function of star formation activity as determined by the strength
of emission lines. They found that the LF of the most passive galaxies
has a shallow faint end slope of $-0.54\pm0.02$ whereas the LF of the
most active star forming galaxies had a steeper slope of
$-1.50\pm0.03$.  Rojas et al. (2003b) find that void galaxies have
fairly strong emission lines and higher star formation rates than wall
galaxies, hence their blueness. However, these void galaxies do not
have average SFRs quite as high as the most extreme Madgwick et al.
sample, thus the void galaxy LF should have a faint end slope 
shallower than $\alpha=-1.5$.

Void galaxies have been studied using CDM simulations plus
semi-analytic models. Mathis \& White extracted mock galaxy catalogs
from high resolution $\Lambda$CDM plus semi-analytic simulations. They
measure the luminosity function as a function of environment,
splitting the catalogs by both mass and volume. They find that there
are no very bright galaxies in voids and that the slopes of the LFs
do not change greatly with environment, although the faint end slopes
are found to be $\alpha \approx -1.4$ in the high density environments
and $\alpha \approx -1.6$ in the low density environments. Both of
these values are steeper than observed although the trend of $\alpha$
is in the same sense as we find in figure \ref{fig:rho}, i.e. LFs
measured from lower density galaxies are steeper than LFs from
galaxies in higher density environments. Mathis \& White point out
that for a void galaxy LF to have a steeper slope, the void region
would have to be populated by faint, dwarf galaxies but it does not
appear that there is a large excess of dwarf galaxies in void regions,
consistent with observations (Bingelli 1989; Kuhn, Hopp \&
Els\"{a}sser 1997; Popescu, Hopp \& Els\"{a}sser 1997). They find that
all types of galaxies in the simulations avoid the same regions and
that no class of galaxy appears to populate the voids defined by
bright galaxies. Benson et al. (2003) have investigated the effects of
photoionization on the shape of the LF. They conclude that feedback,
in the form of supernova-driven winds and photoionization, can
significantly flatten the faint end slope of the LF. Benson et
al. also note the ``hump'' around M=-17 in the LF of smaller mass
halos of ($M< 10^{13}h^{-1}\msun$) and attribute that feature to the
relative contribution of central and satellite galaxies, i.e., at
magnitudes fainter than M=-17, satellite galaxies are more numerous
than central galaxies.

We conclude that galaxies in voids are similar to late-type
galaxies in the low density wall environments. They are fainter,
bluer, diskier and are forming more stars, possibly due to a slower
evolutionary process, but they do not seem to be a different
population, dominated by dwarf galaxies.

\section{Conclusions}
\label{sec:conc}

Using the sample of void galaxies obtained by Rojas et al. (2003a), we
measure the LF from 10$^3$ void galaxies which cover a
wide range of absolute magnitude, $-21.5<$M$_{\rm r}<-14.5$. The void
galaxy LF is well fit by a Schechter function with values of
$\Phi^*$=0.0019 $\pm 0.0004 h^3$Mpc$^{-3}$, M$_{\rm r}^*$ -
5log$h=19.74 \pm0.11$ and $\alpha=-1.18 \pm0.13$.

The fainter value of M$_{\rm r}^*$ is consistent with the result of
Rojas et al. (2003a) who find that void galaxies are fainter than wall
galaxies. The value of $\alpha$ is consistent with values of $\alpha$
found from other analyses of galaxies with similar photometric and
spectroscopic properties. The overall shape of the void galaxy LF is
similar to the shape of the LF of wall galaxies that have similar
photometric properties to the void galaxies. The faint end slope is
shallow compared to that of CDM models without the effects of feedback
taken into consideration. Mathis \& White (2002) conclude that the
galaxies in voids are isolated and are not surrounded by a large
population of dwarf galaxies. 

Once the SDSS is complete, it will cover a wider angle and the need to
use the UZC and SSRS2 to find fainter void galaxies will be
removed. It will then be possible to measure the LF of void galaxies
over a wide range of M$_{\rm r}$ values from one sample alone. We
predict that the final sample will contain 10$^4$ void galaxies, which
will be large enough that the void galaxy sample itself can be split
as a function of color, morphology or star formation rate to
investigate the form of the LF further. We will also be able to look
at the properties of void galaxies as a function of density to see
whether void galaxies with $\delta \rho / \rho < -0.9$ have
different properties or even if such galaxies exist. 

\section*{Acknowledgments} 

Funding for the creation and distribution of the SDSS Archive has been
provided by the Alfred P. Sloan Foundation, the Participating
Institutions, the National Aeronautics and Space Administration, the
National Science Foundation, the U.S. Department of Energy, the
Japanese Monbukagakusho, and the Max Planck Society. The SDSS Web site
is http://www.sdss.org/.

The SDSS is managed by the Astrophysical Research Consortium (ARC) for
the Participating Institutions. The Participating Institutions are The
University of Chicago, Fermilab, the Institute for Advanced Study, the
Japan Participation Group, The Johns Hopkins University, Los Alamos
National Laboratory, the Max-Planck-Institute for Astronomy (MPIA),
the Max-Planck-Institute for Astrophysics (MPA), New Mexico State
University, University of Pittsburgh, Princeton University, the United
States Naval Observatory, and the University of Washington.

MSV acknowledges support from NSF grant AST-0071201 and the John
Templeton Foundation. We thank Peder Norberg, Michael Blanton and
David Goldberg for useful conversations.

\end{document}